\documentclass[a4paper]{jpconf}
\usepackage{graphicx}
\usepackage[verbose]{wrapfig}
\usepackage{amsmath, amssymb}

\begin{document}
\title{Homoclinic chaos in a pair of parametrically-driven coupled SQUIDs}

\author{M.~Agaoglou$^1$, V.~M.~Rothos$^1$, H.~Susanto$^2$}

\address{$^1$
Department of Mechanical Engineering, Faculty of Engineering, Aristotle University of Thessaloniki, Thessaloniki 54124, Greece}

\address{$^2$
Department of Mathematical Sciences, University of Essex, Wivenhoe Park, Colchester CO4 3SQ, United Kingdom}

\ead{rothos@auth.gr}

\begin{abstract}
An rf superconducting quantum interference device (SQUID) consists of a superconducting ring interrupted by a Josephson junction (JJ). When driven by an alternating magnetic field, the induced supercurrents around the ring are determined by the JJ through the celebrated Josephson relations. This system exhibits rich nonlinear behavior, including chaotic effects. We study the dynamics of a pair of parametrically-driven coupled SQUIDs arranged in series. We take advantage of the weak damping that characterizes these systems to perform a multiple-scales analysis and obtain amplitude 
equations, describing the slow dynamics of the system. This picture allows us to expose the existence of homoclinic orbits in the dynamics of the integrable part of the slow equations of motion. Using high-dimensional Melnikov theory, we are able to obtain explicit parameter values for which these orbits persist in the full system, consisting of both Hamiltonian and non-Hamiltonian perturbations, to form so-called Silnikov orbits, indicating a loss of integrability and the existence of chaos.
\end{abstract}



%

\newif\ifamsfonts
\amsfontstrue \ifamsfonts
\font\twlbbb=msbm10 scaled\magstep1 \font\egtbbb=msbm8
\font\sixbbb=msbm6
\newfam\bbbfam
\textfont\bbbfam=\twlbbb \scriptfont\bbbfam=\egtbbb
\scriptscriptfont\bbbfam=\sixbbb

\else
\newcommand{\Bbb}[1]{{\bf#1}}
\fi

\def\Aset{\Bbb{A}}
\def\Hset{\Bbb{H}}
\def\Cset{\Bbb{C}}
\def\Kset{\Bbb{K}}
\def\Nset{\Bbb{N}}
\def\Qset{\Bbb{Q}}
\def\Pset{\Bbb{P}}
\def\Rset{\Bbb{R}}
\def\Sset{\Bbb{S}}
\def\Tset{\Bbb{T}}
\def\Xset{\Bbb{X}}
\def\Yset{\Bbb{Y}}
\def\Zset{\Bbb{Z}}
\def\Dset{\Bbb{D}}
\def\Fset{\Bbb{F}}

\newtheorem{lem}{Lemma}
\newtheorem{rem}{Remark}
\newtheorem{thm}{Theorem}
\newtheorem{prop}{Proposition}
\newtheorem{cor}{Corollary}[section]
\newtheorem{dfn}{Definition}[section]
\newtheorem{nota}{Notation}[section]
\newcommand{\proof}{\noindent{\it Proof. }\ignorespaces}
\newcommand{\qed}{\relax\hfill\mbox{$\Box$}\par\vskip\topsep}
\newcommand{\ie}{{i.e.}}
\renewcommand{\theequation}{\thesection.\arabic{equation}}
\makeatletter\@addtoreset{equation}{section}\makeatother

\newcommand{\dn}{\mathop{\rm dn}\nolimits}
\newcommand{\sn}{\mathop{\rm sn}\nolimits}
\newcommand{\cn}{\mathop{\rm cn}\nolimits}
\newcommand{\ds}{\mathop{\rm ds}\nolimits}
\newcommand{\cs}{\mathop{\rm cs}\nolimits}
\newcommand{\sech}{\mathop{\rm sech}\nolimits}
\newcommand{\cosech}{\mathop{\rm cosech}\nolimits}
\newcommand{\res}{\mathop{\rm res}\nolimits}
\newcommand{\cons}{\mathop{\rm constant}\nolimits}
\newcommand{\diag}{\mathop{\rm diag}\nolimits}
\newcommand{\Spec}{\mathop{\rm Spec}\nolimits}
\newcommand{\trace}{\mathop{\rm trace}\nolimits}
\newcommand{\abs}[1]{\left | #1 \right |}
\newcommand{\norm}[1]{\left | #1 \right |}
\newcommand{\Sint}[3]{{\vphantom{\int}}_{#1}\!\int_{#2}^{#3}}
\newcommand{\Ors}{\mathop{\rm o}\nolimits}
\newcommand{\id}{\,{\rm d}}
\newcommand{\un}{{\rm u}}
\newcommand{\st}{{\rm s}}
\newcommand{\p}{{\rm p}}
\newcommand{\q}{{\rm q}}
\newcommand{\ce}{{\rm c}}
\newcommand{\lef}{{\rm l}}
\newcommand{\iu}{\mskip2mu{\rm i}\mskip1mu}
\newcommand{\Sing}{\mathop{\rm Sing}\nolimits}


\def\rh{\rightarrow}
\def\lgh{\longrightarrow}
\def\Lgh{\Longrightarrow}
\def\t{\tt}
\def\re{\rm e}
\def\pt{\partial}
\def\bd{\begin{displaymath}}
\def\ed{\end{displaymath}}
\def\bqns{\begin{eqnarray*}}
\def\bi{\begin{itemize}}
\def\ei{\end{itemize}}
\def\beq{\begin{quote}}
\def\eeq{\end{quote}}
\def\ben{\begin{enumerate}}
\def\een{\end{enumerate}}
\def\eqns{\end{eqnarray*}}
\def\bq{\begin{equation}}
\def\bqn{\begin{eqnarray}}
\def\eq{\end{equation}}
\def\eqn{\end{eqnarray}}
\def\i{\imath}
\def\j{\jmath}
\def\a{\alpha}
\def\b{\beta}
\def\e{\varepsilon}
\def\l{\lambda}
\def\L{\Lambda}
\def\v{\vec}
\def\m{\mu}
\def\mm{\rm m}
\def\th{\theta}
\def\vt{\vartheta}
\def\Th{\Theta}
\def\g{\gamma}
\def\G{\Gamma}
\def\hr{\tilde r}
\def\k{\rm k}
\def\x{\rm x}
\def\t{\rm t}
\def\r{\rm r}
\def\n{\nu}
\def\d{\delta}
\def\D{\Delta}
\def\o{\omega}
\def\s{\sigma}
\def\S{\Sigma}
\def\O{\Omega}
\def\p{\phi}
\def\vp{\varphi}

\def\k{\kappa}
\def\Y{\Upsilon}
\def\u{\upsilon}
\def\z{\zeta}
\def\rA{\rm A}
\def\ra{\rm a}
\def\rI{\rm I}
\def\rB{\rm B}
\def\rD{\rm D}
\def\rC{\rm C}
\def\rb{\rm b}
\def\rc{\rm c}
\def\rV{\rm V}
\def\rY{\rm Y}
\def\rE{\rm E}
\def\rK{\rm K}
\def\rH{\rm H}
\def\rL{\rm L}
\def\rM{\rm M}
\def\rZ{\rm Z}
\def\rW{\rm W}
\def\w{\rm w}
\def\rU{\rm U}
\def\rS{\rm S}
\def\rP{\rm P}
\def\rQ{\rm Q}
\def\rq{\rm q}
\def\rp{\rm p}
\def\rF{\rm F}
\def\ry{\rm y}
\def\rG{\rm G}
\def\f{\rm f}
\def\rH{\rm H}
\def\h{\rm h}
\def\rT{\rm T}
\def\rd{\triangleright}
\def\di{\displaystyle}

\renewcommand{\theequation}{\thesection.\arabic{equation}}

\newcounter{saveeqn}
\newcommand{\alpheqn}{\setcounter{saveeqn}{\value{equation}}%
\stepcounter{saveeqn}\setcounter{equation}{0}%
\renewcommand{\theequation}{\mbox{\thesection.\arabic{saveeqn}\alph{equation}}}}
\newcommand{\reseteqn}{\setcounter{equation}{\value{saveeqn}}%
\renewcommand{\theequation}{\thesection.\arabic{equation}}}
\renewcommand{\labelenumii}{\Alph{enumi}. \arabic{enumi}}


\section{Introduction}

We consider a series of electrical circuits in a line and we could derive the following nonlinear lattice equation \cite{Tsironis08}, 
\bq \label{e1_0} \ddot f_{n}+{\gamma}\dot
f_{n}+f_{n}+{\hat{\beta}}{\sin}(2\pi
f_{n})-{\l}(f_{n+1}-2f_{n}+f_{n-1})=0 \eq
The coefficient of the nonlinearity, which corresponds to the so-called Josephson critical current, is modulated in time, which can be realized experimentally by modulating the surrounding temperature. As the temperature also influences the damping due to the unpaired electron, additionally we also consider time-periodically modulated dissipation. 
Compared to \cite{Kenig}, we consider softening nonlinearity as opposed to stiffening one, i.e.\ the nonlinearity has different sign. 

Here, we wish to study the possibility of homoclinic chaos near resonances in typical SQUID lattice, which is
described by eq. (\ref{e1}). In particular, we consider a specific dimer for the lattice equation (\ref{e1}) and show that the unperturbed system has a homoclinic orbit in their collective dynamics. Applying the geomertical method  of singular perturbation theory near the resonaces and Melnikov theory of near integrable Hamiltonian system to predict the chaotic behavior near resonances \cite{Haller}. Moreover, we state the theorem for the exisence of multi-homoclinic orbits near resonance following the references \cite{Haller, HallerRothos}.
In Section 2, we perform a multi-scale analysis for the dimer case and obtain amplitude equations to describe the slow  dynamics of the system of dimer SQUID. It is in the slow dynamics that the homoclinic orbits are found. In Section 3, we perform the analytical method for the existence of homoclinic orbits in the perturbed system. In particular we calculate the Energy-difference function using the Melnikov integral evaluated on the homoclinic solutions and applying the singular perturbation theory we study the dynamics near resonances. In Section 4, we illustrate numerically our analytical results for the SQUID model.

\section{Normal mode amplitude equations}

We consider the 1D case of eq (6) in \cite{Tsironis08}

\bq \label{e1} \ddot f_{n}+{\gamma}\dot
f_{n}+f_{n}+{\hat{\beta}}{\sin}(2\pi
f_{n})-{\l}(f_{n+1}-2f_{n}+f_{n-1})=0 \eq
 for n=1,2
and $f_0=f_3=0$
 we obtain \bqn \label{e1a} \ddot
f_{1}+{\gamma}\dot f_{1}+f_{1}+\hat{{\beta}}{\sin}(2\pi
f_{1})-{\l}f_{2}&=&0\cr \ddot f_{2}+{\gamma}\dot
f_{2}+f_{2}+{\hat{\beta}}{\sin}(2\pi f_{2})-{\l}f_{1}&=&0
 \eqn
 We could consider a small parameter
$\gamma_{0}\to{\epsilon}{\hat\gamma _{0}}$, $\hat{\beta}=\beta+f_{p} \cos (\omega _{p} t)$ , $\gamma=\gamma_{0} $ and drive amplitude
$f_{p}=\epsilon h$. The normal mode frequencies for the linear system are: \bq \label{e2}
{\omega}^{2}_{1}=1+2\pi \beta -{\l}, {\omega}^{2}_{2}=1+2\pi \beta+{\l}\eq

We use multiple time scales to express the displacements: \bq
\label{e3} f_{1,2}(t)=\frac{\sqrt{3\epsilon}}{2}(A_{1}(T){\rm
e}^{{i}{\omega}_{1}t}\pm A_{2}(T){\rm
e}^{{i}{\omega}_{2}t}+c.c)+{\epsilon}^{3/2}f^{(1)}_{1,2}(t)+\ldots
\eq

 where $f_1$ is taken with the positive sign and $f_2$ with the
negative sign, slow time $T={\epsilon}t$.  The drive frequency ${\omega}_{p}$ is
related with ${\omega}_{2}$:
${\omega}_{p}=2{
\omega}_{2}+{\epsilon}{\Omega},\quad
{\omega}_{1}={\omega}_{2}+2{\epsilon}{\Omega}_{1}.$ Substituting multiple time scale
expression to the system of $f_1, f_2$ generates secular terms that yield two coupled equations for the complex amplitudes $A_{1,2}$. Express the complex amplitudes using real amplitudes and phases as
\bq \label{e4} A_{1}(T)=a_{1}(T){ \rm
e}^{i(x_{1}(T)+({\Omega /2}-{2\Omega}_{1})T)}, \quad
A_{2}(T)=a_{2}(T){ \rm e}^{i(x_{2}(T)+{\Omega}T/2)} \eq

the real and imaginary parts of the two secular
amplitude equations become:

\bqn \label{e5}
 \frac{{\rm d}a_{1}}{{\rm d}T}&=&
\frac{3\pi^{3}\beta}{2{\omega}_{1}}a_{1}a^{2}_{2}{\sin}(2(x_{2}-x_{1}))+\frac{{\pi\hat
h }}{2{\omega}_{1}}{\sin}(2x_{1})a_{1}-\frac{1}{2}a_{1} \hat{\gamma}_{0}\cr\cr
 \frac{{\rm d}x_{1}}{{\rm d}T}&=&
-\frac{\Omega}{2}+2{\Omega}_{1}-\frac{3\pi^3\beta}{2{\omega}_{1}}(a^{2}_{1}+2a^{2}_{2}+a^{2}_{2}{\cos}(2(x_{2}-x_{1})))+\frac{\pi{\hat
h}}{2{\omega}_{1}}{\cos}(2x_{1})
 \cr\cr
 \frac{{\rm d}a_{2}}{{\rm d}T}&=&
\frac{3\pi^{3}\beta}{2{\omega}_{1}}a_{2}a^{2}_{1}{\sin}(2(x_{1}-x_{2}))+\frac{{\pi\hat
h }}{2{\omega}_{2}}{\sin}(2x_{2})a_{2}-\frac{1}{2}a_{2} \hat{\gamma}_{0} \cr\cr \frac{{\rm
d}x_{2}}{{\rm d}T}&=&
-\frac{\Omega}{2}-\frac{3\pi^{3}\beta}{2{\omega}_{2}}(a^{2}_{2}+2a^{2}_{1}+a^{2}_{1}{\cos}(2(x_{1}-x_{2})))+\frac{\pi{\hat
h}}{2{\omega}_{2}}{\cos}(2x_{2}) \eqn

\subsection{Analytical expressions for homoclinic orbits}

We performed a multi-scale analysis \cite{Kovacic} for the dimer case and obtained amplitude equations to describe the slow  dynamics of the system of dimer SQUID. It is in the slow dynamics that the homoclinic orbits are found. The homoclinic orbits given by
\begin{equation}\label{18a}
B^{h} (T,I)=\frac{2\delta a^2}{q\cosh (2aT)+p}
\end{equation}
\begin{equation}\label{18b}
\theta^{h} (T,I)=\tan^{-1} \Big(\sqrt{\frac{I(\delta -3)+\Omega_{1}}{I(1-\delta) -\Omega_{1}}} \tanh (aT) \Big)
\end{equation}
\begin{equation}\label{18c}
\chi^{h} _{1} (T,I)=\frac{-a(\delta ^{2} -1)}{\sqrt{p^{2} -q^{2}}} \tan ^{-1} [-\sqrt{\frac{p-q}{p+q}}\tanh (aT)] +(\delta I-\frac{\Omega}{4} )T+x_{1} (0)
\end{equation}
\begin{equation}\label{18d}
\phi ^{h} (T,I)=\chi ^{h}_{1} (T,I)-\theta ^{h} (T,I)
\end{equation}

where
$q= I(\delta^{2} -1)+2\delta \Omega_{1} $,
$p=-\Omega_{1} (\delta ^{2} -4\delta +1)-I(\delta ^{3} -6\delta^{2} +7\delta -2)$ and
$a^{2}=\Omega^{2} _{1}-2I\Omega_{1} (\delta -2)-I^{2} (\delta -3)(\delta-1)$.

\begin{wrapfigure}{r} {0.3\textwidth}
\centering
\includegraphics[width=0.3\textwidth]{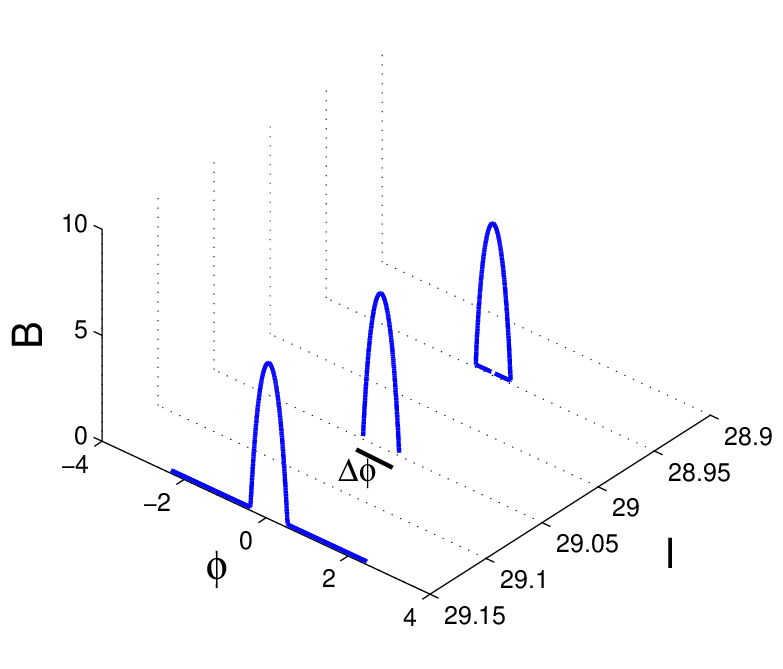}
\caption{Orbits homoclinic to $ \Pi _{0}=\big\{\, (x, y, I, \phi):x=y=0,1<\delta<3,
I>\frac{\Omega_{1}}{3-\delta}\,\big\}$ . For $ I=I_{0}$  (the orbit in the middle),$d\phi/dT =0$ on $ \Pi 
_{0}$, and the orbit is heteroclinic, connecting fixed points on $ \Pi _{0}$ that are $
\Delta\phi$ apart. For $I \lessgtr I_{0}, d\phi/ dT \lessgtr  0$ on $ \Pi _{0}$. The 
parameters are $\delta =1.49704$, $\Omega =-110$, $ \Omega _{1} =27.6085$.} 
\label{fig:fig1}
\end{wrapfigure}

The orbits given by Eqs.\ref{18a}-\ref{18d} are homoclinic to $\Pi_ {0}$. In Fig. 2 we can see some of them. At resonance, for $I=I_{0}=-\Omega/4\delta$, the orbits are heteroclinic, connecting fixed points that are $\Delta \phi$ apart, where $\Delta \phi=\Delta x_{1}-\Delta \theta $. Therefore 
$$\Delta \phi=\frac{2a(\delta^{2} -1)}{\sqrt{p^{2}-q^{2}}} \tanh ^{-1} [\sqrt{\frac{p-q}{p+q}}]-2\tan^{-1} \sqrt{\frac{I_{0}(\delta -3)+\Omega_{1}}{I_{0} (1-\delta)-\Omega_{1}}}$$

\section{Homoclinic intersections in the perturbed system}
In this  section we calculate the Energy-difference function using the Melnikov integral evaluated on the homoclinic solutions. We compute that the energy-difference function is

\bqn
\label{eq14} {\Delta}^{N}{\cal
H}(\phi_{0})&=&\cos (2(\phi_{0} +N\Delta \phi))-\cos (2\phi_{0})+\cr && \frac{\sin\frac{ (N\Delta \phi)}{2}}{\sin \frac{\Delta \phi}{2}} 2 (\sin 2\phi_{0})\sin (N\Delta \phi)-\cr &&\frac{2\xi N}{h\delta}(-2\Delta \phi -\Delta \theta-\frac{\tilde{L}}{I_{0}})
\eqn  
When the dissipation parameter $\xi<\frac{\sin N\Delta\phi (1+2\frac{\sin N\Delta \phi /2}
{\sin \Delta \phi /2})}{\frac{2N}{h\delta} (-2\Delta \phi -\Delta \theta-\frac{\tilde{L}}
{I_{0}})}$ \\then the energy function admits transverse zeros.

\subsection{Dynamics near resonance}

 Now applying the singular perturbation theory we study the dynamics near resonances. The 
equations that describe the dynamics on $M_{\varepsilon}=\big\{\,(x,y,I,\phi): x=x^{\e}(I,
\phi), y=y^{\e}(I,\phi)\,\big\}$ near the resonace $I=I_{0}$ are

\begin{equation}\label{eq33a}
\frac{dI}{dT}=\varepsilon h I \delta \sin (2\phi)-\varepsilon \gamma_{0} I, 
\frac{d\phi}{dT}=-\delta I-\frac{\Omega}{4}+\frac{\varepsilon h\delta}{2}\cos (2\phi)
\end{equation}

According to \cite{Kovacic} and \cite{Haller} in order to study the slow dynamics, which is 
induced by the perturbation on $M_{\varepsilon}$ near resonance, we set a slow variable 
$I=I_{0} + \sqrt{\varepsilon} J$ into Eqs. (\ref{eq33a}) along with a slow time scale $\tau 
=\sqrt{\varepsilon} T$, and we obtain the Hamiltonian system in $(J,\phi)$ 
\begin{equation}
\label{eq35bb}
H(J,\phi)=-\frac{\delta J^{2}}{2} +I_{0} \gamma_{0} \phi +\frac{1}{2} I_{0}h\delta \cos (2\phi)
\end{equation}

\subsection{A homoclinic connection to the sink $p_{\varepsilon}$}
We are in a position to show the existence of an orbit homoclinic to the sink
$p_{\varepsilon}$. Here  we illustrate numerically our analytical results for the SQUID model. In Fig.2a we show that the unperturbed heteroclinic orbit, which asympotes to $p_{c}$ as $T\to \infty$, returns back to a point on the circle of fixed points that is inside the homoclinic separatrix loop. In order to have Silnikov Chaos  \cite{Kovacic} the condition
\begin{equation}\label{eq40}
\phi _{s}<\phi _{c} + \Delta \phi <\phi _{m}
\end{equation} 
has to be satisfied. The $\phi$ values of the fixed points are shown in Fig.2b along with $\phi_{c}+\Delta \phi$ and $\phi_{m}$ for a particular value of $\Omega$. The parameter values for which these $\phi$ values satisfy the condition \ref{eq40} are displayed in Fig.2c.

\begin{figure}[htbp]
\centering
\includegraphics[scale=0.33]{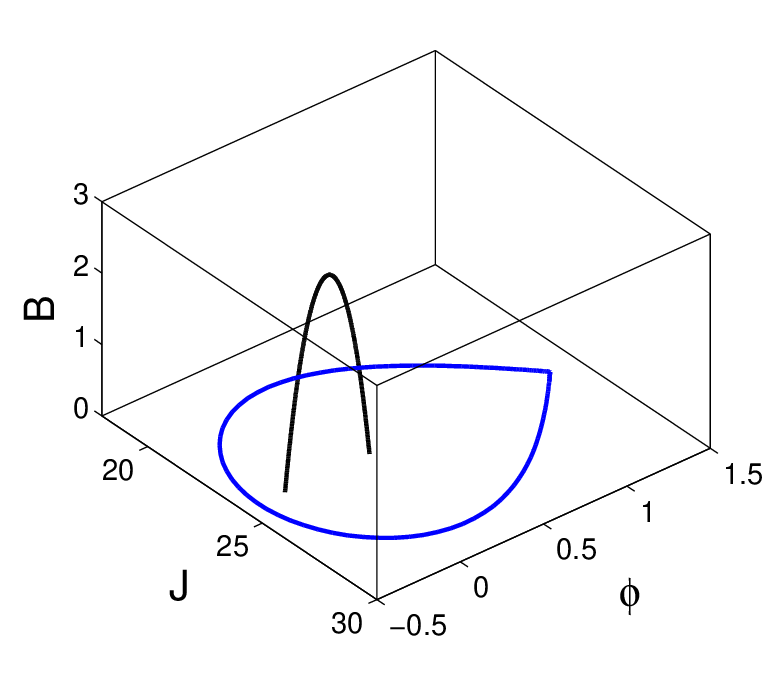}
 \includegraphics[scale=0.33]{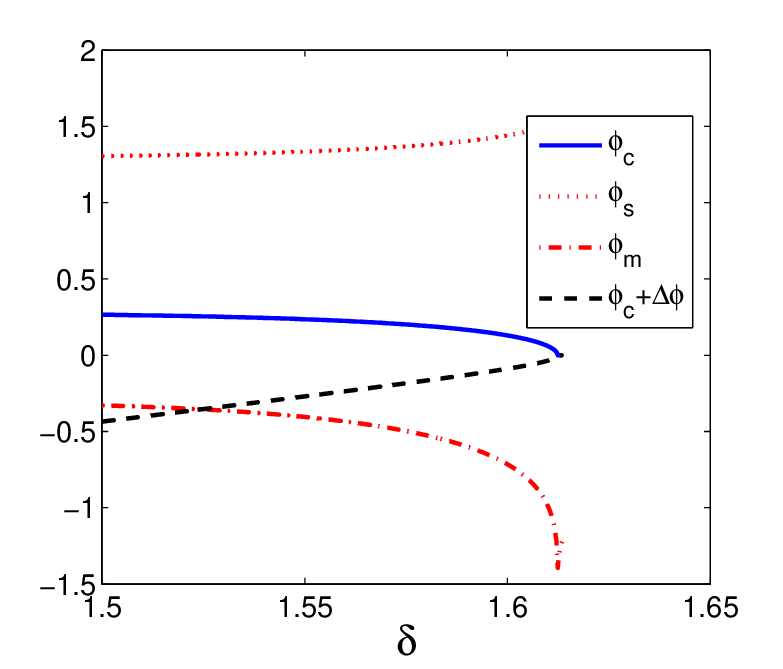}
 \includegraphics[scale=0.33]{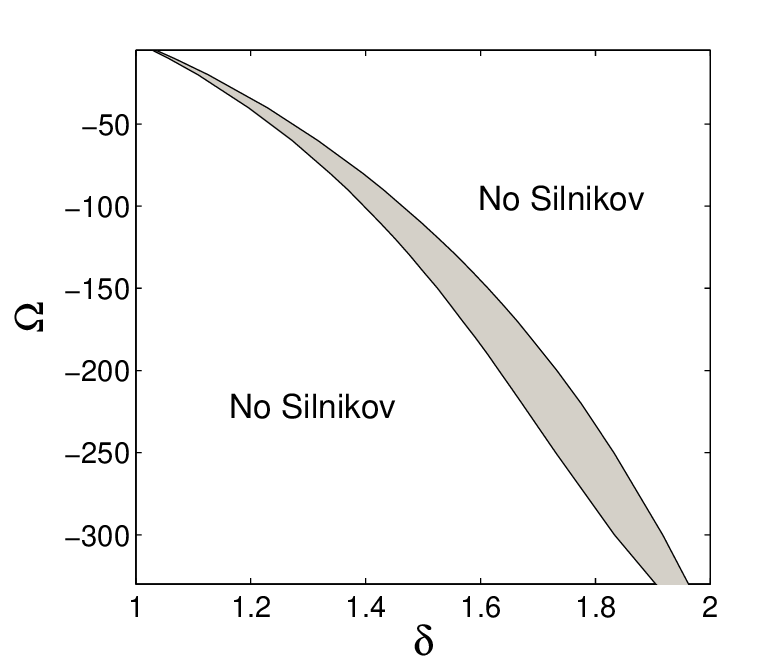}
\caption{a) The heteroclinic orbit given by Eqs. \ref{18a}  and \ref{18d} with $I=I_{0}$, superimposed with the phase portrait of the unperturbed scaled system on $\Pi_{\varepsilon} $ near resonance. The parameters are $\delta=1.55, \Omega=-150, \Omega_{1}=35, \zeta= 0.45, h=1$ ($\xi=\gamma/h$ and $\zeta=\xi/ \delta$). This value of $\zeta$ sets $\phi_{c}=0.233383$ and we can see from the figure that $\phi_{s}<\phi_{c}+\Delta \phi <{\phi}_{m}$.
b) The values of $\phi _{s}, \phi _{c} ,\phi_{c} +\Delta \phi$ and $\phi _{m}$ as functions of $\delta$ , for $\Omega=-150$. For $\delta>1.524$ the condition is satisfied.  c) In gray are indicated the parameter values for which the condition \ref{eq40} is satisfied. In figures b and c: $\varepsilon =0.01$}\label{fig:fig3}
\end{figure} 
\textbf{Summary:}
In this work we studied chaotic dynamics of a pair of parametrically-driven coupled SQUIDs arranged in series and we found under which conditions it can exist by using high dimentional Melnikov theory  \cite{Kovacic}. \\

\small{\textbf{Acknowledgments:}
The work of M.A has been co-financed from resources of the operational program "Education and Lifelong Learning" of the European Social Fund and the National Strategic Reference Framework (NSRF) 2007-2013. The work of V.R. has been co-financed by the European Union (European Social Fund – ESF) and Greek
national funds through the Operational Program “Education and Lifelong Learning” of the National
Strategic Reference Framework (NSRF) – Research Funding Program: THALES – Investing in
knowledge society through the European Social Fund.
}
\section*{References}

\end{document}